\documentclass[prl,aps,twocolumn,superscriptaddress,showpacs]{revtex4}
\usepackage{amsfonts}

\usepackage{dcolumn}
\usepackage{bm}
\usepackage{tikz}
\usepackage[colorlinks=true,citecolor=blue,urlcolor=blue]{hyperref}
\usepackage{amsmath,amsfonts,amssymb,times,natbib}
\usepackage[standard]{ntheorem}
\def\endproof{\vrule height6pt width6pt depth0pt}

\begin{document}


\title{Bell's Nonlocality Can be Tested through Einstein-Podolsky-Rosen Steering}

\author{Jing-Ling Chen}
\email{chenjl@nankai.edu.cn}
 \affiliation{Theoretical Physics Division, Chern Institute of Mathematics, Nankai University,
 Tianjin 300071, People's Republic of China}

\author{Xiang-Jun Ye}
 \affiliation{Theoretical Physics Division, Chern Institute of Mathematics, Nankai University,
 Tianjin 300071, People's Republic of China}
 \affiliation{Key Laboratory of Quantum Information, University of Science and Technology of China, CAS, Hefei 230026, People's Republic of China}


\date{\today}

\begin{abstract}
Quantum nonlocality has recently been classified into three distinct types: quantum entanglement, Einstein-Podolsky-Rosen (EPR) steering, and Bell's nonlocality. Experimentally Bell's nonlocality is usually tested by quantum violation of the Clause-Horne-Shimony-Holt (CHSH) inequality in the two-qubit system. Bell's nonlocality is the strongest type of nonlocality, also due this reason Bell-test experiments have encountered both the locality loophole and the detection loophole for a very long time. As a weaker nonlocality, EPR steering naturally escapes from the locality loophole and is correspondingly easier to be demonstrated without the detection loophole. In this work, we trigger an extraordinary approach to investigate Bell's nonlocality, which is strongly based on the EPR steering. We present a theorem, showing that for any two-qubit state $\tau$, if its mapped state $\rho$ is EPR steerable, then the state $\tau$ must be Bell nonlocal. The result not only pinpoints a deep connection between EPR steering and Bell's nonlocality, but also sheds a new light to realize a loophole-free Bell-test experiment (without the CHSH inequality) through the violation of steering inequality.
\end{abstract}


\maketitle

\emph{Introduction.}---
In 1935, Einstein, Podolsky and Rosen (EPR) indicated that there were some conflicts between quantum mechanics and local realism~\cite{EPR}. If local realism is correct, then quantum mechanics cannot be considered as a complete theory to describe physical reality. In 1964, as a response to the debate raised by the EPR paradox, Bell proposed Bell's inequality by investigating two entangled spin-1/2 systems (i.e., two qubits)~\cite{Bell}. The violation of Bell's inequality, or the violation of local realism, by quantum entangled states implies Bell's nonlocality. This is well-known as Bell's theorem, which has established what quantum theory can tell us about the fundamental features of Nature, and been widely regarded as ``the most profound discovery of science"~\cite{Stapp}. Until now, the fundamental theorem has achieved ubiquitous applications in different quantum information tasks, such as quantum key distribution~\cite{crypt}, communication complexity~\cite{Brukner}, and  random number generation~\cite{Random10}.

As a \emph{no-go theorem} for nonexistence of local hidden variable (LHV) model, Bell's theorem definitely tell us that no physical theory of LHV models can ever reproduce all of the predictions of quantum mechanics. A remarkable point is that Bell's nonlocality can be tested experimentally, instead of a purely philosophical debate. However, the original Bell's inequality is not feasible for experimental test. Consequently, an improved version, the so-called Clause-Horne-Shimony-Holt (CHSH) inequality~\cite{chsh}, was later proposed and verified by experiments~\cite{Freedman}\cite{Aspect}.


Bell's nonlocality concerns measurements made by observers on a pair of particles. Let $\tau_{AB}$ be a two-qubit state shared by Alice and Bob. For some specific measurement settings of the two observers, if  the state $\tau_{AB}$ cannot have a LHV model description, then one says that it possesses a property of  Bell's nonlocality. Experimentally, one usually chooses the state as a maximally entangled state, 
and then detects whether it will violate the CHSH inequality  $\mathcal
{I}_{\rm CHSH}\leq 2$. The maximal violation value by theoretical prediction reads $ \mathcal {I}^{\rm max}_{\rm CHSH}=2 \sqrt{2}\simeq 2.8284$, which is evidently larger than 2. Thus, if an experiment can observe a quantum violation larger than the classical bound, then Bell's nonlocality of the state $\tau_{AB}$ is tested.

However, there are usually two types of loopholes in the Bell test experiments: one is the locality loophole, the other is the detection loophole.
The locality assumption in the EPR paper means that sufficiently distant events cannot change the outcome of a nearby measurement. Nevertheless, an entirely closing of the locality loophole in principle requires that the two particles are in a spacelike separated configuration, this is impossible in a pratical experiment. For the detection loophole, it is usually due to the imperfect detectors and inevitable photon loss during the spatial distribution of entanglement.
The detection loophole can be closed by improving the detector efficiency and some related quantum technologies. Before 2015, some very important experiments have been reported to test Bell's nonlocality~\cite{Ou,Shih,Kwiat,Weihs,Tittel,Rowe,Ansmann,Hofmann,Scheidl,Marshall,Gerhardt,Giustina,Christensen}. But, these experiments have not yet closed both loopholes simultaneously.

Very recently, an experimental loophole-free violation of the CHSH inequality has been reported by using entangled electron spins separated by 1.3 km~\cite{Hensen}. It is the first Bell test experiment, in which physicists  successfully close both the locality and the detection loopholes simultaneously. This ``history-making" experiment not only confirms the ``spooky action at a distance" is an inherent feature of quantum world, but also is useful for
developing the ultrasecure cryptographic devices~\cite{naturenews}. It takes physicists 50 years to achieve a loophole-free Bell test since Bell's discovery (and 80 years since EPR's argument). As another point of view, this fact reflects that testing Bell's nonlocality without loopholes is a long-standing problem.

\begin{figure}[t]
\includegraphics[width=65mm]{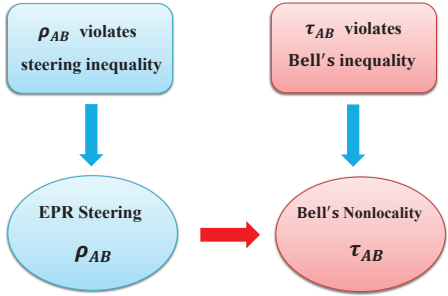}\\
\caption{(Color online) Illustration of testing Bell's nonlocality through EPR steering. Traditionally, Bell's nonlocality of the two-qubit state $\tau_{AB}$ is revealed by its violation of the CHSH inequality. Based on our result, Bell's nonlocality can be tested through EPR steerability of the two-qubit state $\rho_{AB}$, where $\rho_{AB}$ is a mapped state of $\tau_{AB}$ as shown in our theorem.}\label{fig1}
\end{figure}

This gives rise to a natural question: Can Bell's nonlocality be tested in an alternative way, such that the locality loophole can be avoided automatically (and the detection loophole can be closed by the recent mature quantum technologies)? In this Letter, we trigger an extraordinary approach to investigate Bell's nonlocality, which is strongly based on EPR steering. We shall present a theorem, showing that for any two-qubit state $\tau_{AB}$, if its mapped state $\rho_{AB}$ is EPR steerable, then the state $\tau_{AB}$ must be Bell nonlocal. The EPR steerability of the state $\rho_{AB}$ is revealed by quantum violation of EPR steering inequality. Therefore, Bell's nonlocality of the state $\tau_{AB}$ can be tested in a smart way through the violation of EPR steering inequality, instead of using any Bell's inequality. The central idea is illustrated in Fig. \ref{fig1}. Our result not only pinpoints a deep connection between EPR steering and Bell's nonlocality, but also provides a feasible approach to experimentally test Bell's nonlocality without loopholes.

\emph{EPR steering and LHS model.}--- Immediately after the publication of the EPR paper, Schr\"odinger made a response by presenting the notions of ``entanglement" and ``steering".
According to Edwin Schr\"odinger, quantum entanglement is ``the characteristic trait of quantum mechanics" that distinguishes quantum theory from classical theory ~\cite{Schrodinger35}. The notion of ``steering" is closely related to the statement of ``spooky action at a distance", which Einstein was disturbed all the time. Steering is just such a quantum feature that manipulating one object will instantaneously affect another, even it is a far away one.

The research field of steering has been a desert till 2007, when Wiseman, Jones, and Doherty~\cite{WJD07}\cite{WJD07PRA} reformulated the idea and placed it firmly on a rigorous ground.
Since then EPR steering has gained a very rapid development in both theories ~\cite{SU,QKD,AVN,He,Jevtic,Skrzypczyk,Bowles,CS,Kogias,Piani} and experiments ~\cite{NP2010,NC,PRX,NJP, AS12,Schneeloch,Sun,Li}. Most research topics as well as research approaches in the field of Bell's nonlocality have been transplanted similarly to the field of EPR steering. For instance, steering inequalities have been proposed to reveal the EPR steerability of quantum states, very similar to the violation of Bell's inequalities reveals Bell's nonlocality.

\begin{figure}[t]
\includegraphics[width=56mm]{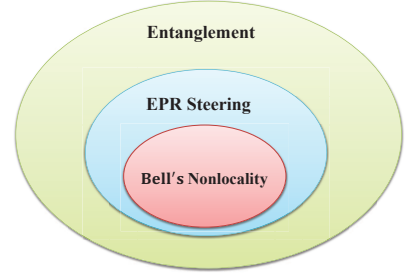}\\
\caption{(Color online) Hierarchical structure of quantum nonlocality. Bell's nonlocality is the strongest type of quantum nonlocality. If a state possesses EPR steerability or Bell's nonlocality, then the state must be entangled. EPR steering is a form of nonlocality intermediate strickly between entanglement and Bell nonlocality. }\label{fig2}
\end{figure}

According to Ref.~\cite{WJD07}, quantum entanglement, EPR steering and Bell's nonlocality are called by a joint name as `` quantum nonlocality", which possesses an interesting hierarchical structure: quantum entanglement is a superset of steering, and Bell's nonlocality is a subset of steering (see Fig.~\ref{fig2}). It is worthy mention that, in 1989 Werner gave a formally mathematical definition for entanglement and proposed the Werner state (or the isotropic state) to show Bell's nonlocality was a subset of quantum entanglement~\cite{werner89}.

Bell's nonlocality is associated with the violation of LHV models. Similarly, EPR steering excludes the possibility of the so-called local hidden state (LHS) models.
In a bipartite steering scenario, Alice prepares a two-qubit state $\rho_{AB}$, she keeps one and sends the other to Bob. She announces that she can remotely ``steer"
Bob's particle into different quantum states simply by measuring her own particle with different settings. To Bob, who does not trust Alice. Bob doubts that she may send him
some nonentangled qubits and fabricate the results using her knowledge about the LHS of his particles. Bob's task is to prove that no such hidden states exist.

 To do this, Bob asks Alice to perform some projective measurements on her qubit and tell him the corresponding  measurement results. Bob's projective measurement is given by
\begin{eqnarray}
\hat{\Pi} ^{\hat{n}_A}_{a}=\frac{\openone + (-1)^a \;\vec{\sigma}\cdot \hat{n}_A}{2}
\end{eqnarray}
where $\hat{n}_A$ is the measure direction, $\openone$ is the $2\times 2$ identity matrix, ${\vec \sigma}=(\sigma_x, \sigma_y, \sigma_z)$ is
the vector of the Pauli matrices, and $a$ is the measurement outcome (with $a=0, 1$). After Alice performs the projective measurement
on her qubit, the state $\rho_{AB}$ collapses to Bob's
conditional states (unnormalized) as
\begin{eqnarray}
\tilde{\rho}^{\hat{n}_A}_a={\rm
tr}_A[(\hat{\Pi} ^{\hat{n}_A}_{a} \otimes \openone) \rho_{AB}], \;\;\; a=0, 1.
\end{eqnarray}
To prove there exists a LHS model for $\rho_{AB}$ is to prove that, for any measurement $\hat{\Pi} ^{\hat{n}_A}_{a}$ and outcome $a$, one can always find a hidden state ensemble $\{ \wp_{\xi}
\rho_{\xi} \}$ and the conditional probabilities $\wp(a|\hat{n},\xi)$, such that the following relation
\begin{eqnarray}\label{LHS1}
&&\tilde{\rho}^{\hat{n}_A}_a=\sum_{\xi} \wp(a|\hat{n}_A,\xi) \wp_{\xi}
\rho_{\xi},
\end{eqnarray}
is always satisfied. Here $\xi$'s are the local hidden variables, $\rho_{\xi}$'s are the hidden states,
$\wp_{\xi}$ and $\wp(a|\hat{n},\xi)$ are probabilities satisfying
$\sum_\xi\wp_{\xi}=1$, and $\sum_a \wp(a|\hat{n}_A,\xi) =1$ for a fixed $\xi$, and
$\rho_B={\rm tr}_A\rho_{AB}=\sum_\xi \wp_{\xi} \rho_{\xi}$ is Bob's reduced density matrix (or Bob's unconditioned state)~\cite{WJD07,WJD07PRA}.
If there exist some specific measurement settings of Alice, such that Eq. (\ref{LHS1}) cannot be satisfied, the one may conclude that the state $\rho_{AB}$ is EPR steerable (in the sense of Alice steers Bob's particle).

Our main result is the following theorem.

\textbf{Theorem}: For any two-qubit state $\tau_{AB}$ shared by Alice and Bob, one can map it into a new state defined by
\begin{eqnarray}\label{rhomu}
\rho_{AB}=\mu \; \tau_{AB} +(1-\mu) \tau'_{AB},
\end{eqnarray}
with $\tau'_{AB}={\tau }_{A}\otimes \openone/2$, ${\tau }_{A}=\textrm{tr}_{B}[\tau_{AB} ]=\textrm{tr}_{B}[\rho_{AB} ]$ being the reduced density matrix at Alice's side,  and $\mu=\frac{1}{\sqrt{3}}$, if $\rho_{AB}$ is EPR steerable, then $\tau_{AB} $ is Bell nonlocal.

\emph{Proof.} The implication of the theorem is that, the EPR steerability of the state $\rho_{AB}$ determines Bell's nonlocality of the state $\tau_{AB} $. Namely, the nonexistence of LHS model for $\rho_{AB}$ implies the nonexistence of LHV model for $\tau_{AB}$. We shall prove the theorem by proving its converse negative proposition: if the state $\tau_{AB} $ has a LHV model description, then the state $\rho_{AB}$ has a LHS model description.

Suppose $\tau_{AB}$ has a LHV model description, then by definition£¬ for any projective measurements $A$ for Alice and $B$ for Bob, one always has the following relation
\begin{eqnarray}\label{LHV}
P(a,b|A, B, \tau )=\sum_{\xi }P(a|A,\xi )P(b|B,\xi )P_{\xi }.
\end{eqnarray}
Here $P(a,b|A,B,\tau )$ is the joint probability, quantum mechanically it is computed as $P(a,b|A,B,\tau )=\textrm{tr}[(\hat{\Pi} ^{\hat{n}_A}_{a}\otimes \hat{\Pi} ^{\hat{n}_B}_{b})\; \tau ]$, $a$ and $b$ are measurement outcomes (with $a, b=0, 1$),  $P(a|A,\xi ), P(b|B,\xi )$ and $P_{\xi }$  denote some (positive, normalized) probability distributions satisfying
\begin{eqnarray}
\sum_{a=0}^{1}P(a|A,\xi )=1, \;\; \sum_{b=0}^{1}P(b|B,\xi )=1, \;\; \sum_{\xi }P_{\xi }=1.
\end{eqnarray}

Let the measurement settings at Bob's side be picked out as ${x, y, z}$. In this situation, Bob's projectors are $\hat{\Pi} ^{x}_{b}$, $\hat{\Pi} ^{y}_{b}$, $\hat{\Pi} ^{z}_{b}$, respectively. Since the state $\tau_{AB}$ has a LHV model description, based on Eq. (\ref{LHV}) we explicitly have
\begin{subequations}  \label{ELHV}
\begin{eqnarray}
P(a,0|A, x, \tau_{AB} )&=&\sum_{\xi }^{}P(a|A,\xi )P(0|x,\xi )P^{}_{\xi },\\
P(a,1|A, x, \tau_{AB} )&=&\sum_{\xi }^{}P(a|A,\xi )P(1|x,\xi )P^{}_{\xi },\\
P(a,0|A, y, \tau_{AB} )&=&\sum_{\xi }^{}P(a|A,\xi )P(0|y,\xi )P^{}_{\xi },\\
P(a,1|A, y, \tau_{AB} )&=&\sum_{\xi }^{}P(a|A,\xi )P(1|y,\xi )P^{}_{\xi },\\
P(a,0|A, z, \tau_{AB} )&=&\sum_{\xi }^{}P(a|A,\xi )P(0|z,\xi )P^{}_{\xi },\\
P(a,1|A, z, \tau_{AB} )&=&\sum_{\xi }^{}P(a|A,\xi )P(1|z,\xi )P^{}_{\xi }.
\end{eqnarray}
\end{subequations}

We now turn to study the EPR steerability of $\rho_{AB}$. Suppose there is a LHS model description for ${\rho }_{AB}$, then it implies that, for Eq. (\ref{LHS1}) one can always
find the solutions of $\{\wp(a|\hat{n}_A,\xi), \wp_{\xi}, \rho_\xi\}$ if Eq. (\ref{ELHV}) is valid. The solutions are given as follows:
\begin{subequations}\label{LHS2}
\begin{eqnarray}
&&\wp(a|\hat{n}_A,\xi)=P(a|A,\xi ), \\
&& \wp_{\xi}={P}_{\xi }, \\
&& \rho_\xi=\frac{\openone+\vec{\sigma}\cdot \vec{{r}}_{\xi }}{2},
\end{eqnarray}
\end{subequations}
where the hidden state $\rho_\xi$ has been parameterized in the Bloch-vector form, with
\begin{eqnarray}
\vec{r}_{\xi }=\mu\; \left({2P(0|x,\xi )-1},{2P(0|y,\xi )-1},{2P(0|z,\xi )-1}\right)
\end{eqnarray}
being the Bloch vector for density matrix of a qubit.
Equation (\ref{ELHV}) is helpful for proving the existence of a LHS model description, and the detail verification is given in Supplementary Materials. This ends the proof. \hfill \endproof

The theorem has some important physical applications.

\emph{Application 1.---} Our theorem directly has a practical application: it can overcome the loopholes, especially the locality loophole, in the Bell test experiments.
 The essence of the theorem is revealing Bell's nonlocality without any Bell's inequality, but with the violation of EPR steering inequality. In the EPR steering test experiment, it does not require two particles are space-like separated because the quantum feature of \emph{steering} just allows measurements at Alice's site affecting Bob's site, thus naturally escapes any locality loophole.

 For example, let us test Bell's nonlocality of the maximally entangled state
\begin{eqnarray}\label{2qubitstate1}
&&|\Psi\rangle=\frac{1}{\sqrt{2}}(|00\rangle+ |11\rangle)
\end{eqnarray}
without the CHSH inequality. Based on our theorem, it is equivalent to test the EPR steerability of the following two-qubit state
\begin{eqnarray}\label{2qubitstate2}
\rho_{AB}= \frac{1}{\sqrt{3}}\; |\Psi\rangle\langle \Psi| +(1-\frac{1}{\sqrt{3}})\; {\tau }_{A}\otimes \frac{\openone}{2},
\end{eqnarray}
with ${\tau }_{A}=\openone/2$. The state (\ref{2qubitstate2}) is nothing but the Werner state with the visibility equals to $1/\sqrt{3}$, its steerability can be tested  by using the EPR-steering inequality
proposed in Ref. \cite{NP2010}
\begin{eqnarray}\label{sn}
\mathcal{S}_{N}=\frac{1}{N}\sum_{k=1}^N\langle
A_k\vec{\sigma}_k^B\rangle\leq C_N
\end{eqnarray}
with $N=6$ as well as $N=10$. Here $\mathcal{S}_{N}$ is the steering parameter for $N$ measurement settings, and $C_N$ is the classical bound, with $C_6=(1 + \sqrt{5})/6\simeq 0.5393$ and $C_{10}\simeq 0.5236$. The maximal quantum violations of the steering inequalities are $\mathcal{S}_{6}^{\rm max}=\mathcal{S}_{10}^{\rm max}=1/\sqrt{3}\simeq 0.5774$, which beat the classical bound. Indeed, the steerability of the Werner state has been experimentally detected in \cite{NP2010} by the steering inequality (\ref{sn}). As stated in \cite{NP2010}: ``\emph{because the degree of correlation required for EPR steering is smaller than that for violation of a Bell inequality, it should be correspondingly easier to demonstrate steering of qubits without making the fair-sampling assumption} [i.e., closing the detection loophole]", thus people have confidence in obtaining a loophole-free test of Bell's nonlocality in the alternative way based on the theorem.

\emph{Application 2.---} The theorem naturally provides a steering-based Bell's nonlocality criterion, which is expressed as: Given an EPR steerable two-qubit state $\rho_{AB}$,  if the matrix
\begin{eqnarray}\label{rhomup}
\tau_{AB}=\frac{1}{\mu} \; \rho_{AB} -(\frac{1}{\mu}-1) \tau'_{AB},
\end{eqnarray}
is a two-qubit density matrix, then $\tau_{AB} $ is Bell nonlocal. Evidently, the criterion depends strongly on the EPR steering criterion for the state $\rho_{AB}$ (unless we have an explicit formula to calculate the EPR steerability, similar to Wootters's concurrence~\cite{Wootters} for entanglement of two qubits). Up to now, there have been two relatively efficient steering criteria: one is the steering inequality as shown in (\ref{sn}), the other one is the \.Zukowski-Dutta-Yin criterion~\cite{Zukowski}.
Combining these two steering criteria and the above Bell's nonlocality criterion, one can detect Bell's nonlocality for a wide class of two-qubit mixed states without any Bell's inequality.

\emph{Conclusion.}--- In conclusion, experimentally Bell's nonlocality of two qubits is usually tested by violation of the CHSH inequality. The locality loophole and the detection loophole are prevalent in the Bell-test experiments. Therefore achieving a loophole-free Bell-test experiment is a long-standing task. EPR steering is a weaker nonlocality in comparison to Bell's inequality, in a steering experiment, the locality assumption is not needed and the detection loophole can be avoided by improving the detector efficiency. In this work, we have presented a theorem, showing that Bell's nonlocality can be tested through EPR steering. The result not only pinpoints a deep connection between EPR steering and Bell's nonlocality, but also sheds a new light to realize a loophole-free Bell-test experiment through the violation of steering inequality. We expect experimental progress in this direction in the near future.

J.L.C. is supported by the National Basic Research Program (973
Program) of China under Grant No.\ 2012CB921900 and the Natural Science Foundations of China
(Grant Nos.\ 11175089 and 11475089).


\end{document}